\documentstyle[12pt,epsfig]{article}

\begin{document}

  \baselineskip=18.6pt plus 0.2pt minus 0.1pt


  \makeatletter
  \@addtoreset{equation}{section}
  \renewcommand{\theequation}{\thesection.\arabic{equation}}
  \begin{titlepage}
  \title{
  \hfill\parbox{4cm} {\normalsize GNPHE/03-04\\hep-th/0303198}\\
  \vspace{1cm}
       {\bf
              M-theory on $G_2$  manifolds  and    the method of  $(p,q)$   
brane webs  }
  }
  \author{
  Adil Belhaj
    {}
  \\[7pt]
{\it \small  National Grouping of High Energy Physics, GNPHE}\\  [1pt]
{\small  and}
  \\[1pt]  {\it \small   Lab/UFR High Energy Physics, Department of 
Physics}\\[1pt]
 {\it \small Faculty of Sciences, Rabat, Morocco}
 {}
{}
  \\[15pt]
{ \it \small   Department of  Mathematics  and Statistics,
  Concordia University}\\[1pt]
{ \it \small  Montr\'eal, Qu\'ebec,  Canada  H4B 1R6}
\\
 }

  \maketitle \thispagestyle{empty}
  \begin{abstract}
Using a reformulation of the  method of $(p,q)$ webs, we study the
four-dimensional  $N=1$ quiver theories from M-theory on seven-dimensional
manifolds with $G_2$ holonomy.  We first  construct such
manifolds  as  $ U(1)$
quotients  of eight-dimensional toric hyper-K\"ahler manifolds, using
$N=4$ supersymmetric  sigma  models.  We show that these geometries,
in general, are given by real cones on $\bf S^2$ bundles
over complex two-dimensional toric varieties, $ \cal \bf V^2=
{{\bf C}^{r+2}/ {{\bf C}^*}^r}$.  Then we discuss the connection
between  the  physics content of M-theory on such  $G_2$ manifolds and
the method of $(p,q)$ webs.  Motivated by a result of Acharya
and  Witten [hep-th/0109152], we reformulate the method of $(p,q)$
webs  and reconsider the derivation of the gauge theories using toric geometry
Mori vectors  of  $ \cal \bf V^2$ and
brane charge constraints.  For ${\bf WP^2}_{w_1,w_2, w_3}$, we
find that the gauge group is given by
$ G=U( w_1n)\times U(w_2n)\times U(w_3n)$. This is
required by the anomaly cancellation condition.

  \end{abstract}
{\tt  KEYWORDS}: M-theory, $G_2$ manifolds, Toric geometry.
  \newpage
  \newpage
  \end{titlepage}
  \newpage
  \def\be{\begin{equation}}
  \def\ee{\end{equation}}
  \def\bea{\begin{eqnarray}}
  \def\eea{\end{eqnarray}}
  \def\nn{\nonumber}
  \def\l{\lambda}
  \def\t{\times}
  \def\[{\bigl[}
  \def\]{\bigr]}
  \def\({\bigl(}
  \def\){\bigr)}
  \def\p{\partial}
  \def\o{\over}
  \def\ta{\tau}
  \def\cm{\cal M}
  \def\R{\bf R}
  \def\b{\beta}
  \def\a{\alpha}
  \newpage
  \section{Introduction}
 Since the   discovery of superstring dualities,   four-dimensional
supersymmetric quantum
  field theories ($QFT_4$)  have been a  subject  of great interest
in connection with superstring compactification on Calabi-Yau manifolds and
D-brane physics \cite{1,2,3,4}. For example,  embedding $N=2$  $QFT_4$ in
type IIA superstring  compactified on Calabi-Yau threefolds,  with K3
fibration, has found a very nice geometric description using the so-called
geometric engineering method \cite{5,6,7,8,9,10}. In this program, these
models, which give  exact results for the moduli space of  the type IIA  
Coulomb
branch, are represented by Dynkin quiver diagrams of  Lie algebras
\cite{7,8,9,10}.\par
 Quite  recently, a special interest has been devoted  to four-dimensional
gauge models preserving only four supercharges \cite{101,102}.  These field  
models
admit a very nice description  in the so-called  $(p,q)$ webs [13-24].
This method concerns the study of $N=1$  four-dimensional quiver
theories  arising on the world-volume of D3-branes  transverse to singular
Calabi-Yau threefolds, $CY^3_B$. The subscript here refers to type IIB string
geometry. The manifolds are complex cones over complex   two-dimensional toric
varieties  $\cal \bf V^2$, e.g. del  Pezzo  surfaces.
They are mirror manifolds of local Calabi-Yau threefolds $CY^3_A$ given
by  elliptic and $C^*$  fibrations over the  complex plane.  Under local
mirror symmetry, a  D3-brane in type IIB geometry  becomes a D6-brane
wrapping a $T^3$  in type IIA mirror geometry. In this way, the $N=1$
four-dimensional quiver theories can be obtained from
D6-branes wrapping 3-cycles $S_i$ in the mirror manifold.  For instance, a
D6-brane on   $T^3$,  whose   homology class is
 \be
 [T^3] = \sum\limits _{i=1}^\ell n_i S_i,
     \ee
 where $\{ S_i, i=1,\ldots,\ell\}$ form a basis of $ H_3(CY^3_A,Z)$,
gives   a   four-dimensional $ N=1$ supersymmetric   gauge theory with
gauge   group
     \be
      G=\prod\limits _{i=1}^{\ell} U(n_i),
     \ee
and   quiver matrix
\be       I _{ij}= S_i\cdot S_j.
     \ee
In  eqs. (1.1) and (1.2),   the  vector  $n_i$ is    specified by the
anomaly cancellation  condition
  \be
      \sum\limits_{i=1}^{\ell} I_{ij}n_i=0.
     \ee
  The above  identities in the method of $(p,q)$ webs  are very exciting.
First, the same  equation  forms   have  been used in the
geometric engineering of  superconformal  models with eight  supercharges.
In this case, the quiver matrix is identified  with an affine $ADE$  Cartan
matrix  $K$ and the gauge group is  $G=\prod_{i}SU (
s_{i}n) $.  The positive integers $s_{i}$ appearing in   $G$ are the usual
Dynkin weights.  They form a special positive definite integer vector $%
{s}=\left( s_{i}\right) $ satisfying ${K}_{ij}s_{j}=0$, as  required by the 
vanishing
 of the beta
 function.  Second, for $\ell=3$ corresponding to complex  two-dimensional
weighted  projective spaces in type IIB geometry,    the   physics content
with unitary gauge groups and charged chiral matter  seems to be similar
to four-dimensional $N=1$  models obtained from  M-theory  on singular
$G_2$ manifolds  studied first in  \cite{22}, see also \cite{23,230}.
These manifolds are constructed as circle quotients of eight-dimensional toric
hyper-K\"ahler (HK) manifolds.  Following \cite{22},   the twistor space
over the  weighted projective space ${\bf WP^2}_{m,m,n}$ has  an
interpretation in type IIA superstring as an intersection of three groups of
D6-branes with multiplicities $m,m,n$ leading to $SU(m)\t  SU(m)\t SU(n)$
gauge symmetry.   According to this  feature, one might  ask the
following question. Is there a connection between the  approach of $(p,q)$
webs and M-theory on $G_2$ manifolds\footnote{ Besides this similarity,
much of the D6-branes physics content can be interpreted in M-theory on
local $G_2$ manifolds, leading to $N=1$ supersymmetric models in four
dimensions.}?  However,  this connection may naturally lead  to the need
of a reformulation of the method of $(p,q)$ webs. The reason for this
is that the gauge symmetry  in the M-theory compactification involves
the weights of the weighted projective space  $\bf WP^2$.
 In   this paper we address this question using toric geometry data of
$G_2$ manifolds as  $U(1)$ quotients of eight-dimensional HK manifolds,
and by reconsidering the method of $(p,q)$ webs. This study may complete
the analysis of \cite{24}   dealing with   discrete
$G_2$ orbifolds     using the McKay correspondence \cite{25}.\\
 Our program will  proceed   in two steps:\\
(i) We study $G_2$  manifolds as $U(1)$
  quotients of eight-dimensional toric HK manifolds,  $ X_7=
{X_8/ U(1)}$. The manifold $X_8$  is  obtained  using relevant
constraint  equations in terms of two-dimensional $N=4$  sigma-models
with $U(1)^r$ gauge symmetry and $r+2$ hypermultiplets \cite{22,23,26}.
We show that the resulting seven-dimensional  manifolds,  in general,
are given by  real cones on   ${\bf S^2}$ bundles
over complex two-dimensional toric varieties
 \begin
  {equation} {{\cal \bf V}^2} = {{\bf C}^{r+2}/ {{\bf C}^*}^r}.
  \end{equation}
 Explicit  models   are  presented
in terms  of two-dimensional  $N=2$ sigma model  realizations
of  $ \cal \bf V^2$. \\
(ii)  We discuss the link between  the physics content of M-theory on
such $G_2$ manifolds and the methods of $(p,q)$  webs.  In particular,
we reconsider and   reformulate  the $(p,q)$ web    equations
using the toric geometry  Mori vectors  of $ {{\cal \bf V}^2}$
and set of brane  charge constraint equations. For the  weighted projective
space ${\bf WP^2}_{w_1,w_2,w_3}$, for example, we find   the following
gauge    group
\be
      G=U( w_1n)\times U(w_2n)\times U(w_3n).
\ee
This is required by the anomaly cancellation  condition. With an  appropriate 
choice of weight vectors,
we recover the result of Acharya and Witten given in \cite{22}. \par
 The plan of
 this paper is as follows.  In  section 2, we  briefly review the main
lines of toric
geometry  method for  treating   complex manifolds. Then we give the
interplay between  the toric geometry and two-dimensional $N=2$
supersymmetric gauge theories.   In section 3, we
  study  $G_2$ manifolds as $ U(1)$   quotients of
   eight-dimensional toric HK manifolds $X_8$  constructed  from
$D$-flatness conditions of two-dimensional field theory with  $N=4$  
supersymmetric.  Then we identify the  $U(1)$ symmetry group with   the toric
  geometry circle actions  of $X_8$ to   present  quotients $X_7=
  {X_8/ U(1)}$ of $G_2$  holonomy.  Explicit  models   are given
  in  terms of real  cones  on an  $ {\bf S^2}$ bundle over  complex 
two-dimensional
  toric varieties $ \cal \bf V^2$. In section 4,  we engineer
$N=1$ quiver models from $G_2$ manifolds. We discuss the link
between  the physics content of M-theory on such $G_2$ manifolds and
the method  of $(p,q)$  webs.  We reconsider and
reformulate  the $(p,q)$ equations using the toric geometry  Mori vectors  of 
$ {{\cal \bf V}^2}$  and set of brane charge constraint equations.
In particular,  for the  weighted projective space
${\bf WP^2}_{w_1,w_2,w_3}$, we find that the   gauge group  is given by (1.6).
        In section 5, we give illustrating  applications.
         In
  section 6,   we give our conclusion.
  \section{     Toric geometry}

  In  this section,  we  collect a  few facts  on  toric geometry
 of complex manifolds. These facts   are  needed  later to   construct
a special  type of
    $G_2$ manifolds,  as  $ U(1)$  quotients of eight-dimensional
 toric HK  manifolds.  Roughly speaking, toric manifolds are complex
  $n$-dimensional manifolds
   with $T^n$ fibration over $n$-dimensional  base spaces with boundary
  \cite{7,10,27,28,29,30}.
    They exhibit
    toric actions $U(1)^n$  allowing us   to encode the
  geometric properties of the complex spaces in terms of simple
  combinatorial data of polytopes ${\Delta}_n$ of the $R^n$ space.
  In this correspondence, fixed points of the toric actions $U(1)^n$
  are associated with the vertices of the polytope ${\Delta}_n$, the
  edges are fixed one-dimensional lines of a subgroup $U(1)^{n-1}$
  of the toric action $U(1)^n$, and so on.  Geometrically, this means
  that  the $T^n$ fibers   can degenerate over the boundary of the
  base.  Note that in the case where the base space is compact, the
  resulting toric  manifold  will be compact as well.\\
 In string theory,  the power  of the toric geometry  representation
 is due to the following  points: \\
   (1) The  toric data of the  polytope  ${\Delta}_n$ have similar features
 to  the $ADE$  Dynkin diagrams leading to non-abelian  gauge symmetries in
type II superstring compactifications on Calabi-Yau manifolds \cite{7,8,9,10}.
   (2)  The toric fixed   loci, which correspond to
  the vanishing cycles, have been known to be associated with
D-brane  charges \cite{28}.  The latter   will be  used  in section 4
 to discuss    the physics content
  of M-theory on our   proposed manifolds of $ G_2$  holonomy, using  a
 reformulation of the method of  $(p,q)$ webs in type II superstring on
 Calabi-Yau threefolds.\par
   To illustrate the main idea of  toric geometry, let us
  describe  the philosophy of this subject  through certain useful  examples.\\
   {\it (i) $\bf P^1$  projective space}.\\
   This is the simplest example in toric geometry which turns out to
   play a crucial role in  the  building
   blocks  of higher-dimensional toric
varieties  and  in the study of the small resolution of $ADE$
singularities of  local  Calabi-Yau manifolds.
 $\bf P^1$ has an
  $U(1)$ toric action
 \be  z\to e^{i \theta}z\ee with  two fixed points
  $v_1$ and $v_2$ on the real line. The latter  points,  which  can
  be generally chosen as $v_1=-1 $ and $v_2=1$, describe
  respectively north and south poles of the real two sphere $\bf S^2
  \sim \bf  P^1$. The corresponding one-dimensional polytope is just
  the segment $[v_1,v_2]$ joining the two points $v_1$ and $v_2$.
  Thus, $\bf P^1$ can be viewed as a segment $[ v_1,v_2]$  with a
  circle on top,  where the circle vanishes at the end points $v_1$
  and $v_2$.
  \\ {\it (ii) $\bf P^2$ projective space.}\\
      $\bf P^2$ is a complex  two-dimensional toric variety defined by
  \begin {equation}
  {\bf P^2}= {C^{3}\setminus\{(0,0,0)\}\over C^*},
  \end{equation}
   where   $ C^*$  acts as follows
     \begin {equation}
   (z_1,z_2,  z_{3}) \to ( \lambda z_1,\lambda z_2,\lambda z_{3}).
\end{equation}
 It  admits   an  $U(1)^2$ toric action \bea  (z_1,z_2,  z_{3}) \to ( e^{i
  \theta_1} z_1,e^{i \theta_2} z_2, z_{3}), \eea exhibiting three
  fixed points $v_1$, $v_2$ and $v_3$. The corresponding polytope
  ${\Delta}_2$ is a finite sublattice of the $ \bf Z^2$ square lattice.
  It describes the intersection of three $\bf P^1$'s   defining a
  triangle ($v_1 v_2 v_3$) in the $ \bf R^2$ plane. A convenient choice
  of the data of these  three vertices  is as follows : $ v_1= (1,0)$,
  $v_2=(0,1)$, and $ v_3=(-1,-1)$. Thus,  ${\Delta}_2$ has three edges,
  namely $[v_1,v_2]$, $[v_2,v_3]$ and $[v_3,v_1]$ stable under the
  three $U(1)$ subgroups of $U(1)^2$; two subgroups are just the two
  $U(1)$ factors,  while the third subgroup is the diagonal one.
 $\bf P^2$ can be viewed as a triangle over
  each point of which there is an elliptic curve  $ T^2$. This torus
 shrinks to a
  circle at each segment   $[v_i,v_j]$  and  it shrinks to a  point at
 each $ v_i$.  The above  toric realization can be pushed further for
describing the same phenomenon involving complex  $n$-dimensional toric
 varieties  that are more complicated than projective spaces.  The latter
 spaces  can   be expressed in the following form
  \begin {equation}
  {{\cal \bf V}^n} = {C^{n+r}\setminus U\over {C^*}^r},
  \end{equation}
  where  now we have $r$ ${C^*}$ actions  given by
  \begin {equation}
   {C^*}^r: z_i   \to \lambda ^{Q_i^a} z_i,\quad  i=1,2,\ldots, n+r; \quad a=1,
  2,\ldots,r.
  \end{equation}
    In this equation,  $Q_{i}^{a}$ are integers. For each $a$ they form
the so-called  Mori
  vectors in toric  geometry. They   generalize  the  weight vector
$(w_i)$ of the  complex  $n$-dimensional  weighted projective space
 ${\bf WP}^n_{w_1,\ldots,w_{n+1}}$.  $U$ is a subset  of
  $C^{k}$ chosen by triangulation \cite{7}.
  \\
 Eq. (2.5) means  that  ${{\cal \bf V}^n}$  has a   $ T^n$  fibration,
 obtained
    by dividing   $ T^{n+r}$  by the $U(1)^r$  gauge symmetry
    \be
    z_i\to e^{iQ^a_i \vartheta^a} z_i,\quad a=1,\ldots,r,
    \ee
   where $\vartheta^a$ are the generators of the $ U(1)$ factors.
$\cal \bf  V^n$  can
 be  represented by a  toric
     diagram $ \Delta({{\cal \bf V}^n})$ spanned by  $ k=n+r$
     vertices $ v_i$ of an  $\bf Z^n$ lattice satisfying
      \begin {equation}
    \sum \limits _{i=1}^{n+r} Q_i^a v_ i=0,\quad a=1,\ldots,r.
  \end{equation}
   The  toric  geometry manifolds we have been describing  have  an
 interesting
  realization   through    linear sigma models,    where one considers
  two-dimensional   supersymmetric $N=2$  gauge systems  with $U(1)^r$
 gauge
  group  and $n+r$    chiral  fields $X_i$  with  a
 $ Q_i^a $ matrix
  charge \cite{31}.  In this way, the K\"ahler manifold   ${{\cal \bf V}^n}$
 is  the minimum  of the
  $D$-term potential ($D^a=0$), up to $U(1)^r$ gauge transformations,
  namely
       \begin {equation}
    \sum \limits _{i=1}^{n+r} Q_i^a |X_i|^2=R_a,
  \end{equation}
  where  the $ R_a$'s are Fayet-Iliopoulos (FI) coupling parameters.
  The (local) Calabi-Yau condition  is satisfied by
  \begin {equation}
    \sum \limits _{i=1}^{n+r} Q_i^a=0,  \quad \forall a,
  \end{equation}
   which means that the  system  flows in the infra-red to a non-trivial
  superconformal
  theory \cite{31,32}.
 Under local  mirror symmetry,  this toric  Calabi-Yau sigma model maps
 to Landau-Ginsburg (LG) models \cite{33,34,35,36}.  In this  way,
 the mirror version  of the constraint equation  (2.9),   giving the LG
 superpotential,  reads
\be
\sum\limits_i y_i=0
\ee
subject to
\be
\prod \limits_i y_i^{Q_i^a}=e^{-t_a},
\ee
 where $ y_i$ are  LG  dual  chiral fields which can be related, up some
field changes, to sigma model fields,
and where  $t_a$'s are the complexified FI parameters defining  now the
complex deformations  of  the LG Calabi-Yau superpotentials.\\
\\
  Note that  the  above two-dimensional  $N=2$ toric sigma   models can be
 extended to  $ N=4$ sypersymmetry models  with hypermultiplets leading to
 toric  HK geometries \cite{37}.  In the rest of this paper, we will use
toric geometry and HK   analysis  to study seven-dimensional   manifolds
  with
 $G_2$ holonomy. The latter  are  $U(1)$  quotients of  eight-dimensional
toric HK  manifolds $X_8$.

 \section{  $G_2$ manifolds as  $ U(1)$  quotients}
        \subsection{ $G_2$ manifolds  and  $N=4$ $D$-flatness conditions }
 It is known that in order to get a semi-realistic four-dimensional theory
 from M-theory  it is necessary to consider a compactification  on
 a seven-dimensional  manifold $ X_7$  with $G_2$ holonomy  [42-51].
 In this way, the resulting   models with $N=1$ supersymmetry  depend
on the geometric properties of $X_7$. For instance, if $X_7$ is smooth,
 the low-energy theory contains, in addition  to $ N=1$ supergravity,
 only abelian gauge symmetry and no charged chiral fermions. Non-abelain
 gauge  symmetries can be obtained by considering limits  where $X_7$
develops  $ADE$ orbifold  singularities using wrapped  M2-branes on
vanishing 2-cycles \cite{39}.  However, the presence of  conical
 singularities  leads  to  charged chiral  fermions.
 Following \cite{22}, an interesting analysis for building such
geometries is to consider quotients of
 eight-dimensional toric  HK  manifolds $X_8$ by an $U(1)$   circle symmetry.
  The $ U(1)$  group has been  chosen such  that it commutes with the  $SU(2)$
 symmetry,  permuting the three  complex structures of HK geometries.
 A priori,    there are  many ways  to choose the $ U(1)$  group  action.
 Two situations   have been given in \cite{22} but here we will identify
 the $ U(1)$  group  with the  toric  geometry  circle action  of complex
 subvarieties
  within  HK geometries. In  particular, we will  use  the HK analysis
 to present  explicit models  with $G_2$ holonomy group leading to
interesting $N=1$ supersymmetric gauge theories in four dimensions.
 To do so,   we consider  two-dimensional   $N=4$
supersymmetric gauge theories  with   $U(1)^r$  gauge symmetries   and
 $r+2$ hypermultiplets with  a  $Q^a_i$  matrix charge \cite{32,37}.   The
 $N=4$  $D$-flatness
   equations  of such models   are  generally given by
   \be
    \sum\limits_{i=1}^{r+2} Q^a_i\[\phi_i^{\alpha}{\bar \phi}_{i \beta}
           +\phi_{i \beta}
          {\bar \phi}_i^{\alpha}\]=\vec \xi_a
          \vec\sigma^\alpha_\beta, \quad  a=1,\ldots,r.
          \ee
   In these equations,
            $\phi_i^{\alpha}$'s
          denote   $r+2$   component field doublets of
          hypermultiplets, $\vec \xi_a$ are  $r$ FI 3-vector
 couplings  rotated by $SU(2)$
          symmetry, and
           $\vec \sigma^\alpha_\beta$ are the
          traceless $2\t2$ Pauli matrices.
             In
  this construction,
             for  each
  $U(1)$ factor, there are three real constraint equations
  transforming as  an iso-triplet of $SU(2)$ $R$-symmetry ($SU(2)_R$)
  acting on the HK structures.\\ Using the $SU(2)_R$ transformations
  \be \phi^{\alpha}=\varepsilon^{\alpha\beta}\phi_ 
{\beta},\quad\overline{(\phi^\alpha)}=\overline{\phi}_\alpha,\quad 
  \varepsilon_{12}=\varepsilon^{21}=1,
   \ee
 and  replacing the Pauli matrices by their expressions, the identities
  (3.1) can be
   split as
  follows  \bea \sum\limits_{i=1}^k Q_i^a(
  |\phi^1_i|^2-|\phi^2_i|^2) &= &\xi^3 _a\\ \sum\limits_{i=1}^kQ_i^a
  \phi^1_i \overline{\phi}_{i}^2&=&\xi^1_a+i{\xi^2}_a
  \\
  \sum\limits_{i=1}^k Q_i^a\phi^2_i
  \overline{\phi}_{i}^1&=&\xi^1_a-i{\xi^2}_a. \eea Dividing the
  resulting  space of  (3.3-5) by  $U(1)^r$ gauge transformations,
  we find precisely an
eight-dimensional toric HK manifold   $X_8$.   However,
 explicit  solutions of these geometries  depend on the values of the FI
   couplings.  Taking   $\xi^1_a=\xi^2_a=0$  and
  $\xi^3_a >0$,  (3.3-5) describe the
  cotangent bundle over  complex  two-dimensional toric varieties \cite{26}.  
Indeed, if 
we set all $\phi^2_i=0$,   we get a  complex  two-dimensional toric variety  
$\cal \bf V^2$  defined by $
  \sum\limits_{i=1}^{2+r} Q_i^a|\phi^1_i|^2 = \xi^3 _a, ( 
a=1,\ldots,r)$.
 Equations (3.4-5) mean
  that the $\phi^2_i$'s define the cotangent orthogonal  fiber
  directions over ${\cal \bf  V^2}$.  This   manifold 
         has  four toric geometry  circle actions:
        $U(1)^2_{base}\times U(1)^2_{fiber}$.   Two  of them   correspond
to the
  ${\cal \bf V}^2$  toric
      base space
       denoted by  $U(1)^2_{base}$,
      while  the remaining ones, $U(1)^2_{fiber}$,  act on  the  fiber
    orthogonal   cotangent directions.
      To get the   corresponding seven-dimensional manifolds with $G_2$
    holonomy, we will
       identify   the $U(1)$ group symmetry  of the quotient used in
\cite{22}  with
         one  finite circle toric action.  Identifying this  $ U(1)$
  symmetry  with one  $U(1)_{fiber}$, one finds the
 following
         seven-dimensional  manifold
  \be
  X_7={X_8/ U(1)_{fiber}}.
   \ee
       Since ${{\bf C^2}/ U(1)}
={\bf R}\times
       {\bf  S^2}$,  this  quotient  space  is now
 isomorphic to an ${\bf R}\times {\bf S^2}$ bundle over  a  ${\cal \bf V}^2$.
   Similarly to \cite{22},  equation (3.6)  describes   real  cones  on
 a  $ {\bf S^2}$ bundle
   over ${\cal \bf  V}^2$.
  Mathematically,  it is not easy to  reveal  that these quotient  spaces
have  $G_2$  holonomy  group. However, one can show this using
 a physical argument. Indeed,  ${\cal \bf  V}^2$, with $ h^{1,0}=h^{2,0}=0$,
 preserves $1/4$ of initial supercharges  and in the
 presence of $ {\bf  S^2}$ it  should be  $1/8$. In this  way,
 the supersymmetry tells us that  the holonomy of (3.6)
 is the $G_2$  Lie group.
 Thus,   M-theory on  the above    seven-dimensional   manifold leads to
  $ N=1$ theory in four dimensions.\\

   \subsection{  Explicit models from  ${\cal \bf V}^2$ geometries}
  
To better  understand the structure
 of (3.3-6),  let us
  give illustrating
models.  In particular  we will consider   special  models  corresponding to
$N=4$ sigma model with conformal invariance.  For  this reason, we will   
restrict ourselves
 to     eight-dimensional  toric HK  manifolds $X_8$ with the Calabi-Yau
 condition (2.10)
 in   $N=4$ supersymmetric  analysis.
   In this way,  the geometry of  $X_8$
 depends on the manner we  choose   the  $U(1)^r$   matrix  gauge charge
 $Q_i^a$ satisfying  the Calabi-Yau condition.    We first study
  complex  two-dimensional weighted
    projective  spaces  $\bf WP^2$, after which we will consider
      the Hirzebruch surfaces. Other extended   models   are also presented.
   \subsubsection {${\cal \bf  V}^2$ as weighted projective spaces}
      For  constructing  these models,   we  consider  an  $U(1)$  gauge
 symmetry
  with three
     hypermultiplets  $\phi_i$ of charges  $(Q_1,Q_2,Q_3)$ such that $
  Q_1+Q_2+Q_3=0$.  One  way  to solve this constraint equation is   to  take $
  Q_1=m_1$, $Q_2=  -m_1-m_2$  and $Q_3= m_2$.  This gives  ${\bf 
WP^2}_{m_1,m_1+m_2,m_2}$  as a base
geometry in the $G_2$  manifold.
          Using examples,  let us  see how we obtain this  geometry. \\
\\
             { \bf Example 1: $(m_1,m_2)=(1,1)$ }. This example
 corresponds to
  three
     hypermultiplets $\phi_i$ with  the     vector charge $ Q_i= (1,-2,1)$. 
   After permuting
  the role of $\phi^1_2$ and
   $\overline{\phi^2}_{2}$  and  making the
  following  field changes  $\phi^1_{1}=\varphi_1$, $
  \phi^2_{1}=\psi_1$ $\phi^1_{3}=\varphi_2$, $
  \phi^2_{3}=\psi_3$, $ - {\overline \phi^2_{2}}=\varphi_2$, $
  {\overline \phi^1_{2}}=\psi_2$, 
eqs. (3.3-5)  become
   \bea
   (|\varphi_{1}|^2+|\varphi_{3}|^2+2|\varphi_{2}|^2)-(|\psi_{1}
   |^2+|\psi_{3}|
   ^2+2|\psi_2|^2)
   &=&\xi^3 \\
  \varphi_{1}\overline {\psi_{1}}+\varphi_{3}\overline
  {\psi_{3}}+2\varphi_{2}\overline {\psi_{2}} &=&0
  \\\overline{ \varphi_{1}} {\psi_{1}}+\overline
  {\varphi_{3}}\psi_{3}+2\overline {\varphi_{2}}\psi_{2}&=&0.  \eea These 
equations    describe a cotangent bundle over
 ${\bf WP^2}_{1,2,1}$.   Indeed,   taking  $\psi_{1}=\psi_{2}=\psi_3=0$,
       eq. (3.7) reduces to $ |\varphi_{1}|^2+|\varphi _{3}|^2+2|\varphi_{2}|^2
   =\xi^3$   and defines    a  ${\bf WP^2}_{1,2,1}$ weighted
    projective
    space,  where  $\xi^3$
  is a K\"ahler real parameter controlling its  size.   Eqs.
  (3.7-9), for generic values  of $\psi_{i}$, can be interpreted to
  mean that  $\psi_{i}$  parameterize  the  orthogonal fiber
  directions on  ${\bf WP^2}_{1,2,1}$.  Dividing by one
  finite  toric geometry  fiber circle action, we  find  a real cone on an
  $\bf S^2$  bundle over ${ \bf WP^2}_{1,2,1}$ with  $G_2$ holonomy.\\
\\ {\bf
  Example 2: $(m_1,m_2)=(1,2)$}. As another example,   we consider a
  vector charge   as  follows $Q_i=(1, -3, 2)$. This example is
  quite similar to the first
    one,  and its treatment  will be  parallel to the first one. 
    After  making similar field changes,  this example describes
 ${\bf
  WP^2}_{1,3,2}$ in the  base geometry of an eight-dimensional manifold.
    After the $U(1)$ quotient,  the corresponding seven-dimensional
 manifold
    $X_7$  will be     a real cone  on   $\bf S^2$ bundle over
 $ {\bf WP^2}_{1,3,2}$. We will
 see  later that this  geometry   leads to a four-dimensional model
 which might  be  related to   the  grand unified  symmetry.
   \subsubsection{${\cal \bf V}^2$ as   Hirzebruch  surfaces  ${\bf F}_n$      
   } ${\bf F}_n$   are complex  two-dimensional toric
 surfaces  defined by non-trivial fibrations of
 a $\bf  P^1$ over  a $ \bf P^1$. These may be viewed as the
 compactification of  complex  line bundles  over $\bf P^1$  by adding
 a point to each fiber at infinity. Such line bundles  are classified by
an  integer $n$, being the first Chern class integrated over $\bf P^1$.  For 
simplicity,  we will restrict ourselves to
   ${\bf F}_0$  with a  trivial fibration.
 A  way  to  write down  the $\bf F_0$ $N=4$   sigma model  is  to 
   start with   one  $\bf P^1$  and then  extend
   the result to ${\bf F}_0$. Indeed, 
    one $\bf P^1$   corresponds to an $U(1)$  two-dimensional  $N=4$
  linear sigma model with two hypermultiplets  with  a vector charge
  $(1,-1)$.    Making a similar analysis of previous examples, the
   $D$-flatness conditions (3.1)  reduce to \bea
   ( |\varphi_1|^2+|\varphi_2|^2) -( |\psi_1|^2+|\psi_2|^2)&=
  &\xi^3 \\
   \varphi_1
  \overline{\psi}_1+\varphi_2 \overline{\psi}_{2}&=&0
  \\
 {\psi}_1 \overline{\varphi_1} +{\psi}_2\overline{\varphi_2} &=&0.
   \eea
   and describe  the  cotangent bundle over  a $\bf P^1$,  defined by $
    |\varphi_1|^2+|\varphi_2|^2=\xi^3$.  
    The model  corresponding to  ${\bf F}_0$  is obtained  by  considering 
  an  $U(1)^2$   two-dimensional  $N=4$
linear sigma model with four  hypermultiplets   with  the following  charges
  \be
      Q_i^{(1)}=(1,-1,0,0), \quad 
           Q_i^{(2)}=(0,0,1,-1).\ee  In this way, $N=4$ $D$-flatness   
constraint equations 
 describe  the  cotangent bundle over  ${\bf F}_0$. After dividing by one
 finite toric  geometry circle action, we get a real cone on $\bf S^2$ bundle
 over ${\bf F}_0$.
\subsection{  Other  models   from $\bf WP^2$}
  Here,
  we study some  extended models    using more
  general $N=4$  two-dimensional gauge theories.
In particular,  we consider
  two possible   generalizations for  $\bf WP^2$.  The first model
 describes   the  blowing  up  of $\bf WP^2$   at one  point. It has
a  similar feature as $ {\bf F}_2$   geometry.   The second  model deals with 
model with  $ADE$  Cartan matrix gauge charges 
leading  to  $ADE$  intersecting geometries.
   \subsubsection{  Blowing  up  of ${\bf WP^2}$  at one point }
 For  simplicity,  we consider  ${\bf WP^2}_{1,2,1}$ as an example.
  This space has   a   $Z_2$ orbifold singularity
      corresponding to  non-trivial fixed points under the  homogeneous
      identification
      \be
      (z_1,z_2,z_3)\equiv(\lambda z_1,\lambda^2 z_2,\lambda z_3).
      \ee
  Taking    $\lambda=-1$,    ${\bf WP^2}_{1,2,1}$  has  a $Z_2$ orbifold
 singularity at $(z_1, z_2,
  z_3)=(0,1,0)$.  This singularity  may be blown up by introducing an
 exceptional divisor.  In  two-dimensional $N=2$  sigma model,  this  can be
  deformed  by introducing   an extra chiral
  field $X_4$ and an $U(1)$ gauge group factor. In this way,  the 
corresponding eight-dimensional  manifolds
 can be
 described by an $U(1)^2$ linear  sigma model with four hypermultiplets with
the following  charges
           \be
            Q_i^{(1)}=(1,-2,1,0)\qquad
           Q_i^{(2)}=(0,-1,0,1).\ee 
 This model  gives  the same
$G_2$ manifold   corresponding to  the   ${\bf F}_2$ Hirzebruch surface.
   \subsubsection{ $ADE$  intersecting geometry }
   Another  generalization    is to consider the
   intersecting  weighted  projective  spaces according to  $ADE$
 Dynkin diagrams by imitating the analysis of $N=2$ sigma model.  This 
involves   two-dimensional    $N=4$ supersymmetric $U(1)^r$ gauge theory with 
$(r+2)$
  $\phi_i^{\alpha}$  hypermultiplets  with  $ADE$   Cartan matrices as  matrix 
gauge charges.
   For simplicity, let us consider the $ A_r$  Lie  algebra where the
   matrix charge is given by $ 
Q^a_i=-2\delta^a_i+\delta^a_{i-1}+\delta^a_{i+1}, 
  a=1,\ldots,r$. Putting  these   equations into the $D$-flatness equations 
(3.1), one
  gets the following  system of $3r$ equations
   \bea
   (|\phi^1_{a-1}|^2+|\phi^1_{a+1}|^2-2|\phi^1_{a}|^2)-(|\phi^2_{a-1}|
   ^2+|\phi^2_{a+1}|^2-
   2|\phi^2_a|^2)&=&\xi_a \\
  \phi^1_{a-1}\overline{\phi^2}_{a-1}+\phi^1_{a+1}\overline{\phi^2}
  _{a+1}-2\phi^1_{a} \overline{\phi^2}_{a}&=&0
  \\
  \phi^2_{a-1}\overline{\phi^1}_{a-1}+\phi^2_{a+1}\overline{\phi^1}
  _{a+1}-2\phi^2_{a} \overline{\phi^1}_{a}&=&0 . \eea   An examination
 of these equations  reveals   that $\cal \bf V^2$
    consists of
   $r$ intersecting
    ${\bf WP^2}_{1,2,1}$ according to the $A_r$ Dynkin  diagram \cite{26}.
 Actually,  this geometry   generalizes the usual $ ADE$  geometry
corresponding to  two-cycles  of K3 surfaces  \cite{7,8,9,10}.     One expects 
to have a  similar  feature  in  the compactification of  M-theory on  $G_2$  
manifolds with intersecting ${\bf WP^2}_{1,2,1}$'s.\par 
 The previous  analysis is  also possible    for   models with del Pezzo 
   surfaces  as a  base geometry of $G_2$ manifolds.  Note that,
 these  surfaces    have been used   in the  building   of  $N=1$  
supersymmetric gauge
 theories in four dimensions using the so-called $(p,q)$   webs.   These  
gauge theories
arise  on the world-volume of D3-branes  transverse to local Calabi-Yau 
threefolds $CY^3_B$
 given by complex cones over  del Pezzo  surfaces \cite{12,19}.    In this 
present work, we will show that this physics   is  related to  
 of M-theory on
 $G_2$ manifolds  with  two complex dimension toric manifolds  in the base 
geometry. 
 \section{ On  M-theory on  $G_2$  Manifolds   and $(p,q)$ webs  }
 So far, we have constructed a special type of $ G_2$ manifolds  as  $U(1)$   
quotients of
 eight-dimensional  toric HK  manifolds.
   This section will be concerned  with M-theory on  such  manifolds.   We 
will try to find  a
 superstring interpretation of this using D-brane physics. In particular, we 
will discuss the
connection between the physics content of  M-theory on  such $G_2$  manifolds 
and the method of
 $(p,q)$  webs, leading to $N=1$ supersymmetric gauge theories in four 
dimensions.  The analysis
 we will be using here is based on a
reconsideration of the method of $(p,q)$ webs   and reformulating
 the intersection number structures in terms of toric geometry  data of $\cal 
\bf V^2$  varieties. \par  Before  proceeding,  let us     recall quote  some 
crucial  points  supporting 
our discussion.
 On  one hand,  according to  \cite{22},   M-theory on $G_2$ manifolds  as $ 
U(1)$  
quotient  of
 eight-dimensional conical toric  HK   manifolds,  has an interpretation in 
terms of intersecting
D6-branes in type  IIA superstring model.  In particular, the
geometry of $ {\bf WP^2}$
  describes  the intersection of
   three sets   of D6-branes.  For example the geometry
     $ {\bf WP^2}_{m,m,n}$ with  $m$, $m$ and $n$ relatively prime
     corresponds to a pair of two spheres of   $A_m$ singularities
     and a single of  $A_n$ singularities  two spheres. In type
     IIA superstring picture,  this is equivalent to the intersection of
   three sets   of D6-branes with multiplicities $m$, $m$ and $n$ leading to  
 $ SU(m)\times SU(m)\times SU(n)$ gauge symmetry  with chiral multiplets in the
  $(m,{\bar m},1)+(1,m,{\bar n})+({\bar m},1,n)$ bi-fundamental
  representations.  This gauge system   is   represented  by  a    quiver 
    triangle
which may be viewed as the toric geometry graph of $ {\bf WP^2}$.
 On the other hand,   the same  physics content  models   can  appear  in type
  IIA superstring compactified on  local elliptic
 fibration  Calabi-Yau threefolds $CY^3_A$  in  the presence of
D6-branes
   wrapping      3-cycles $S_i$  and   filling the four-dimensional Minkowski
    space-time \cite{101,102}.
   In terms of gauge theory, each 3-cycle $ S_i$ is associated to a
  single gauge group  factor and the intersection numbers   (1.4) 
  count the number of $N=1$
   chiral multiplets which transform in the bi-fundamental
   representation.  We will see later that the information of  gauge  system 
is encoded
    in the  intersection numbers.    Under mirror  symmetry, the physics 
of D6-branes
   wrapping 3-cycles  maps to
IIB D3-branes transverse to local Calabi-Yau threefolds $CY^3_A$.
The latter are    complex cones  over
 del Pezzo surfaces or more  generally complex two-dimensional toric manifolds
   $\cal \bf V^2$.
  In this way, the $ N=1$ four-dimensional  quiver theory can be obtained from
   type IIB geometry using the so called  $( p,q)$  brane webs. Indeed,   the
 toric skeletons of these  varieties are defined by  the  $( p,q)$ web charges 
of
  D5-branes. They
correspond  to the loci of points at which some 1-cycles of the elliptic  
curve  fibration
  of $\cal
 \bf V^2$  shrinks  to zero radius \cite{28}.   The physics
   content of these
 models   can be  determined explicitly from the geometry of the  $(p,q)$ webs.
  More details on this method, see   \cite{12,13}.
  In particular,  if
the vanishing 1-cycles  of the elliptic fibration
   $C_i\equiv(p_i,q_i)$, the intersection  numbers,   in type IIA  mirror 
geometry,    read as
 \be
  I_{ij}=  C_i\cdot C_j=p_iq_j-p_jq_i
\ee
 The set of the ranks  of the gauge groups  $n_i$  is a
   null vector of this matrix, i.e

  \be
      \sum\limits_{i} (p_iq_j-p_jq_i)n_i=0.
     \ee
\par Until this level, the  connection between  the method  of  $(p,q)$   
brane webs  
and the
 Acharya-Witten
   model \cite{22} is not obvious.  We propose that this connection  
  requires the introduction of  the Mori vector 
charge $Q_i^a$ in the
description  of $(p,q)$  webs. Our  solution   was  inspired by the following:\\
(1) The  study of  M-theory on local geometry of  Calabi-Yau threefolds  
having  toric realization in terms of  $(p,q)$ D5-branes of
 type IIB superstrings \cite{28}. In this way, the  D-brane
 charges are associated  with the vanishing cycles in the toric 
representation.\\ (2)  The  result of  Acharya-Witten  on M-theory on
$G_2$ manifolds, where the set of ranks of gauge groups  coincide
with   the weight  vector of the $\bf WP^2$ \cite{22}. \\ (3) The
local mirror symmetry  application in type II superstrings, where
the mirror constraint equations   involve the toric geometry data
of the original manifolds \cite{33,34,35,36}.\\
 Besides  these points,  a  close examination of  the formulation of the  
$(p,q)$   webs
   reveals, however,  that
 the  matrix  intersection (4.1) appears in the ordinary and weighted 
projective
 spaces. Moreover, it does not  carry  any  transparent toric geometry    data
distinguishing  these geometries.
 Taking into  account this observation,  the connection we are after  leads  
us to reformulate the
  intersection  number structures    by introducing  the  toric geometry Mori  
vectors $Q_i^a$ and
 a set of brane charge constraint equations.
To make connection with   \cite{22},   we  restrict ourselves to the   
weighted projective
spaces  where  $\vec{Q^1}=(w_1,w_2,w_3)$.  Given a set of  charges 
$(p_i,q_i)$, $i=1,2,3$,   we propose the  intersection  number formula
  \be
   {\cal I}_{ij} =w_iw_j (p_iq_j-p_iq_j)
   \ee
 with the following constraint equations
 \bea
   w_1^2 p_1+w_2^2p_2+w_3^2p_3=0\nn\\
    w_1^2q_1+w_2^2q_2+w_3^2q_3=0.
    \eea
 Now, the set of ranks  of the gauge groups  $n_i$  should  satisfy the 
following  constraint 
 \be
   \sum\limits_i {{\cal I}}_{ij}n_i=0,
     \ee
 as required by  the anomaly cancellation condition \cite{12,13}.
Using equation (4.4),  it is easy to  see that   this condition
can be  satisfied in terms of the weights of $\bf WP^2$ as follows
\be
     n_i=w_in,
     \ee
and so the  corresponding gauge symmetry is given by \be
      G=\prod\limits_{i=1} U(w_in).
     \ee
 Our   reformulation of the $(p,q)$  webs  has the following nice features:\\
(1) This  formulation  is quite similar to the geometric engineering of
 four-dimensional
 $N=2$ superconformal field theories with gauge  group
 $ G=\prod\limits_{i=1}SU(s_in)$  where
   the $s_i$'s are the usual Dynkin labels  being a  null vector of
 affine Cartan matrices as
 required by the vanishing of the beta function.
 \\(2)  For  $w_i=1$, we recover the  simple model with gauge group $U(n)^3$  
and  matter
 triplication in each bifundamental   \cite{17,18}. \\  (3)
 For $ n=1$, the corresponding   gauge theory  is now
quite similar  to the interpretation
 of M-theory on $G_2$ manifolds given  in \cite{22}.
In this way, the gauge group  reads
\be
      G=\prod\limits_{i=1}^3 U(w_i).
     \ee
 In the infra-red limit the $U(1)$ factors decouple and one is left with the 
gauge  symmetry
\be
      G=\prod\limits_{i=1}^3SU(w_i).
     \ee
Taking  an appropriate  choice of weights, we recover
the physical model given
in \cite{22}.\\
(4) The corresponding field models  is  represented by a  triangle quiver 
diagram
\vspace{0.5cm}
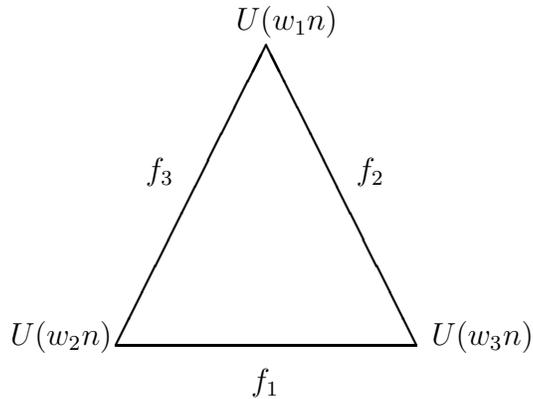
$$
 \mbox{
  \begin{picture}(100,132)(0,0)
  \unitlength=2cm
  \thicklines
   \put(0,0){\line(1,2){1}}
   \put(0,0){\line(1,0){2}}
   \put(2,0){\line(-1,2){1}}
   \put(0.8,2.1){$ U(w_1 n)$}
   \put(-0.7,0){$U(w_2 n)$}
   \put(2.1,0){$U(w_3 n)$}
   \put(0.2,1.1){$f_3$}
   \put(0.9,-0.3){$f_1$}
   \put(1.6,1.1){$f_2$}
 \end{picture}
  }
  \label{four}
$$

\vspace{0.5cm}

 Summarizing,   many aspects of the physics  of  M-theory  on such $G_2$
 manifolds are reproduced by this formulation  of   
type IIB superstring on
  complex lines on  $ \cal \bf  V^2$ in  the  presence of $(p,q)$ brane  
charges, suggesting equivalence of the two descriptions.  Performing
  local mirror  symmetry, we  end with local elliptic type IIA geometry with 
D6-branes  filling
 four-dimensional space-time.  The corresponding gauge group and  quiver 
diagram can be
 obtained using  the  $(p,q)$ webs  toric geometry data of $ \cal \bf  V^2$.
  \section{  Illustrating Models}
In this section we   will  give  two   illustrating applications. They concern 
the examples
 studied in section 3:   ${\bf WP^2 }_{1,2,1}$ and  ${\bf WP^2}_{1,3,2}$.

  \subsection{$ U(n)^2 \times  U(2n)$ gauge  theory}
   Consider,  first,   the geometry of   ${\bf WP^2}_{1,2,1}$ in M-theory 
compactifications.   In $(p,q)$ webs, this  is equivalent to  taking three 
stacks of  branes  each, wrapping  the following  1 cycles
  \be
  C_1=(-2,0), \quad C_2=(0,-1), \quad C_3=(2,4).
  \ee
 In this case,  the intersection numbers read as
\bea
{\cal I}_{12}&=&4\nn\\
{\cal I}_{31}&=&8 \\
{\cal I }_{23}&=&4.\nn
\eea
  For one D6-brane, this   example leads to a  $N=1$    spectrum with gauge 
group  $U(1)^2 \times  U(2)$ gauge group and bifundamental matter.
 This model  agrees  with  the  result of  Acharya and Witten given   in 
\cite{22}.  While for  $ n$ D6-branes, the above  charge configurations    
gives a   $N=1$    spectrum with gauge group  $U(n)^2 \times  U(2n)$ gauge
symmetry  and bifundamental matter.

  \subsection{$ U(n) \times U(2n)\times U(3n)$ gauge model}
 The geometry of  ${\bf WP^2 }_{1,3,2}$ is  very exciting    in  this analysis
  because it may  lead  to the  symmetry of  the grand unified  theory (GUT) 
\footnote{ This  is not a  surprise  since
    it was shown in \cite{47} that one gets the GUT  gauge symmetry from  
M-theory on  $G_2$ manifolds with $A_4$ singularity}.
  For this example,   we    consider  three stacks of $n$ D6-branes  each, 
wrapping  the following  1 cycles
  \be
  C_1=(4,9), \quad C_2=(-1,0), \quad C_3=(0,-1).
  \ee
 In this case,  the  intersection numbers  read as
\bea
{\cal I}_{12}&=&18\nn\\
{\cal I}_{31}&=&12\\
{\cal I }_{23}&=&6\nn
\eea
 This   yields a  $N=1$    spectrum with gauge group $ U(n) \times U(2n) 
\times  U(3n)$ gauge group and bifundamental matter. For $n=1$, one gets U(1)$ 
\times$ U(2)$\times$ U(3)  as gauge symmetry. \par
  Concluding   this section, it is interesting to   make a comment
 regarding the numbers appearing in (5.2) and (5.4),  counting the  number of  
$N=1$ chiral  multiplets $f_i$ in the corresponding gauge systems.  The latter 
have  a remarkable feature which has a  nice interpretation using  the recent 
derivation of local   mirror symmetry in two-dimensional field theory with
$N=2$ supersymmetry \cite{35}. Indeed, in the above two  examples,  $f_i$ can 
be written as follows
\bea
f_1&=&w_2w_3d \nn\\
f_2&=&w_1w_3d\\
f_3&=&w_1w_2d\nn
\eea
where $d$  is   the degree of the following  homogeneous  LG  Calabi-Yau 
superpotentials
\bea
y^{2}+x^{4} +z^{4} +e^txyz&=&0\nn\\
y^{2}+x^{3} +z^{6} +e^txyz&=&0
\eea
 mirror to type IIB  $N=2$ sigma model on the anti-canonical line bundles  over
 ${\bf WP^2}_{1,2,1}$ and  ${\bf WP^2}_{1,2,3}$  respectively.
 \section{  Conclusion}
   In this paper, we have studied  $N=1$   supersymmetric   gauge theories
embedded in  M-theory  on local
  seven-dimensional
   manifolds  with  $G_2$  holonomy  group.   We have engineered  the $N=1$ 
quiver models from
   $G_2$ manifolds,   as   $ U(1)$  quotients of eight-dimensional  toric HK   
manifolds.  The corresponding quiver models have been obtained  using a 
reformulation of the method of $(p,q)$ webs. Our main results may be 
summarized as follows:\\
(i) Using  two-dimensional $N=4$  sigma-models  with $U(1)^r$ gauge symmetry 
and $r+2$ hypermultiplets,  we have constructed a special kind of  $G_2$  
manifolds. The    latter  are      $U(1)$
  quotients of eight-dimensional  toric  (HK)  manifolds,  $ X_7=
{X_8\over U(1)}$. We have  shown  that these seven-dimensional   manifolds,  
in general,
are given by  real cones on   $\bf S^2$ bundles over    complex  
two-dimensional   toric varieties
$ {{\cal \bf V}^2} = {{\bf C}^{r+2}/{{\bf C}^*}^r}$.
 Explicit  models  have been given  in terms  of  $N=2$ sigma model  
realizations  of  $ \cal \bf V^2$. \\
(ii)  We have  discussed  the link between  the physics content of M-theory on 
such $G_2$ manifolds and the method of $(p,q)$  webs.    We have  reconsidered 
 and    reformulated  the method of   the $(p,q)$  webs  using the toric 
geometry  Mori vectors of $ {{\cal \bf V}^2}$  and brane charge constraint
 equations. For the  weighted projective space ${\bf WP^2}_{w_1,w_2,w_3}$,  we 
have  found  that the corresponding  field model  has $
      G=U( w_1n)\times U(w_2n)\times U(w_3n)$ as gauge symmetry group. This is
required by the anomaly cancellation condition.
  \\
  \\ \\ {\bf Acknowledgments}\\
A. Belhaj  would like  to thank  Department of  Mathematics  and Statistics, 
Concordia University, Montreal,   for its kind  hospitality
     during the preparation of this work.  He is   very grateful to   J. McKay 
    for the  invitation, discussions and
     encouragement.
  He would like to thank J. Rasmussen for comments on the manuscript,
      and   H. Kisilevsky, E.H. Saidi  and A. Sebbar  for discussions and
scientific help. 


\newpage
      \begin{thebibliography}{99}


\bibitem{1} S. Kachru, C. Vafa, {\em Exact Results for N=2
   Compactifications of Heterotic Strings}, Nucl. Phys. { \bf B450}
 (1995) 69, {\tt hep-th/9505105}.
 \bibitem{2}
        A. Klemm, W. Lerche, P. Mayr, {\em  K3--Fibrations and Heterotic-Type 
II String Duality},
 Phys. Lett. {\bf B357} (1995) 313, {\tt hep-th/9506112}.
\bibitem{3}
 P.S. Aspinwall, M. Gross, {\em  Heterotic-Heterotic String Duality and 
Multiple K3 Fibrations},
 Phys. Lett. {\bf  B382} (1996) 81, {\tt hep-th/9602118}.
\bibitem{4}
 P.S. Aspinwall, {\em  Enhanced Gauge Symmetries and Calabi-Yau Threefolds}, 
Phys. Lett. {\bf  B371}
(1996) 231, {\tt hep-th/9511171}.

 \bibitem{5}
       S. Katz, A. Klemm, C. Vafa, {\em  Geometric Engineering of Quantum 
Field Theories}, Nucl. Phys. {\bf B497} (1997) 173,
 {\tt hep-th/9609239}.

\bibitem{6}
      S. Katz, C. Vafa, {\em Matter From Geometry},
        Nucl. Phys. {\bf B497} (1997) 146, {\tt hep-th/9606086}.

  \bibitem{7} S. Katz, P. Mayr, C. Vafa, {\em Mirror symmetry and exact
  solution of 4d  N=2 gauge theories I}, Adv. Theor. Math. Phys.
  {\bf 1}(1998) 53.
  \bibitem{8} P. Mayr, {\em  Geometric construction of  N=2  of  Gauge 
Theories}, Fortsch.
 Phys. {\bf 47} (1999) 39.
  \bibitem{9} A. Belhaj, A.E. Fallah, E.H. Saidi, {\em On the non-simply
  mirror geometries in type II strings}, {  Classical Quantum Gravity } {\bf 
17}
  (2000) 515.
  \bibitem{10}
   A. Belhaj,  E.H. Saidi, {\em  Toric Geometry, Enhanced non
  Simply Laced Gauge Symmetries in Superstrings and F-theory
  Compactifications}, {\tt hep-th/0012131}.
\bibitem{101}
 M. Cvetic, G. Shiu, A.M. Uranga, {\em  Chiral Four-Dimensional N=1 
Supersymmetric Type IIA Orientifolds from Intersecting D6-Branes},
 Nucl. Phys. {\bf  B615}  (2001) 3, {\tt hep-th/0107166}.

\bibitem{102}
 M. Cvetic, G. Shiu, A.M. Uranga, {\em  Three-Family Supersymmetric 
Standard-like Models from Intersecting Brane Worlds},
Phys. Rev. Lett. {\bf  87} (2001) 201801, {\tt  hep-th/0107143}.

\bibitem{11}
  C.E. Beasley, M. R. Plesser, {\em Toric Duality Is Seiberg Duality}, { JHEP} 
{\bf  0112} (2001) 001, {\tt  hep-th/010905}.

 \bibitem{12}
         A. Hanany, A. Iqbal, {\em Quiver Theories from D6-branes via Mirror 
Symmetry},
  JHEP {\bf  0204} (2002) 009, {\tt  hep-th/010813}.

\bibitem{13}
 B. Feng, A. Hanany, Y-H. He, A. Iqbal, {\em  Quiver theories, soliton spectra 
and
Picard-Lefschetz transformations}, {\tt  hep-th/0206152}.

\bibitem{130}
 S. Franco, A. Hanany, {\em Geometric dualities in 4d field theories and their 
5d interpretation},
 {\tt hep-th/0207006}.

 \bibitem{14}
 S. Franco, A. Hanany, {\em Toric Duality, Seiberg Duality and 
Picard-Lefschetz Transformations},
 {\tt hep-th/0212299}.
       \bibitem{15}
 B. Feng, S. Franco, A. Hanany, Y-H. He, {\em  Symmetries of Toric Duality}, 
JHEP {\bf  0212} (2002)
 076, {\tt  hep-th/0205144}.
\bibitem{16}
B. Feng, A. Hanany, Y-H. He, A.M. Uranga, {\em Toric Duality as Seiberg 
Duality and Brane Diamonds},
 { JHEP} {\bf  0112} (2001) 035, {\tt  hep-th/0109063}.
\bibitem{17}  A.M. Uranga, {\em  Chiral four-dimensional string 
compactifications with intersecting
 D-branes}, {\tt  hep-th/0301032}.

\bibitem{18}
 A.M. Uranga, {\em  Local models for intersecting brane worlds},  JHEP {\bf  
0212} (2002) 058, {\tt
 hep-th/0208014}.
  \bibitem{19}

  D. Cremades, L.E. Ibanez, F. Marchesano, {\em More about the Standard Model 
at Intersecting Branes},
 { \tt  hep-ph/0212048}.



 \bibitem{20}
     A. Hanany, J. Walcher, {\em  On Duality Walls in String Theory},
{\tt hep-th/0301231}.

\bibitem{21}
 D. Cremades, L.E. Ibanez, F. Marchesano, {\em Yukawa couplings in 
intersecting D-brane models},
 {\tt hep-th/0302105}.

 \bibitem{22}
  B. Acharya, E. Witten, {\em Chiral Fermions from Manifolds of
  $G_2$ Holonomy}, {\tt hep-th/0109152}.

 \bibitem{23}
       P. Berglund, A. Brandhuber, {\em Matter From G(2) Manifolds},  Nucl. 
Phys. {\bf  B641}
 (2002) 351, {\tt  hep-th/0205184}.

\bibitem{230}
       S. Gukov, D. Tong, {\em D-Brane Probes of Special Holonomy Manifolds}, 
JHEP {\bf 0204} (2002) 050, {\tt hep-th/0202126 }.

\bibitem{24}
        Y.-H. He, {\em  $G_2$  Quivers}, JHEP {\bf 0302} (2003) 023,  {\tt 
hep-th/0210127}.

\bibitem{25}
J. McKay, {\em  Graphs, Singularities, and Finite Groups},  Proc Symp. Pure. 
Math. {\bf Vo37}
 (1980) 183.
\bibitem{26}
A. Belhaj, {\em  Manifolds of $ G_2$  Holonomy from N=4 Sigma Model},
         J. Phys. {\bf A35} (2002) 8903, {\tt  hep-th/0201155}.

\bibitem{27}W. Fulton, {\em Introduction to Toric varieties},  Annals of Math. 
Studies, No. {\bf  131}, Princeton University  Press, 1993.
  \bibitem{28} N.C. Leung and C. Vafa, Adv. Theor. Math. Phys. {\bf 2} (1998) 
91, {\tt hep-th/9711013}.

  \bibitem{29}
  D. Cox, {\em The homogeneous coordinate ring of a toric variety},
  J. Alg. geom. {\bf 4} (1995) 17.
  \bibitem{30}
  M. Kreuzer, H. Skarke, {\em   Reflexive polyhedra, weights and
  toric Calabi-Yau
   fibrations},
  {\tt  math.AG/0001106}.\\ A.C. Avram, M. Kreuzer, M. Mandelberg,
  H. Skarke, {\em The web of Calabi-Yau hypersurfaces in
   toric varieties},   Nucl. Phys. {\bf  B505} (1997) 625, {\tt  
hep-th/9703003}.\\
  M. Kreuzer, H. Skarke, {\em  Calabi-Yau 4-folds and toric
  fibrations}, J. Geom. Phys. {\bf 26} (1998) 272, {\tt hep-th/9701175}.

  \bibitem{31}  E. Witten, Nucl. Phys. {\bf B403} (1993) 159, {\tt
  hep-th/9301042}.
  \bibitem{32}
     A. Belhaj,  E.H. Saidi, {\em Hyper-K\"ahler Singularities in
  Superstrings Compactification and 2d N=4 Conformal Field Theory},  Classical 
Quantum Ggravity,
  {\bf  18} (2001) 57, {\tt hep-th/0002205}.\\  A. Belhaj, E.H. Saidi, {\em On
   Hyper-K\"ahler  Singularities};
    Mod. Phys. Lett. A, { Vol. 15},  {\bf  No. 29} (2000) 1767, {\tt
    hep-th/0007143}.
  \bibitem{33} K. Hori, C. Vafa, {\em Mirror Symmetry},
  {\tt hep-th/0002222}.
  \bibitem{34} K. Hori, A. Iqbal, C. Vafa, {\em D-Branes And Mirror Symmetry}, 
{\tt hep-th/0005247}.
    \bibitem{35} M. Aganagic, C. Vafa, {\em Mirror Symmetry, D-branes and
     Counting Holomorphic Discs}, {\tt hep-th/0012041}.\\
       M. Aganagic, C. Vafa,
  {\em Mirror Symmetry and  $G_2$ Flop}, {\tt hep-th/0105225}.\\
   M. Aganagic, A. Klemm, C. Vafa, {\em Disk Instantons,
  Mirror Symmetry and the Duality Web}, {\tt  hep-th/0105045}.
  \bibitem{36}
   A. Belhaj, {\em  Mirror symmetry and Landau-Ginzburg Calabi-Yau 
superpotentials  in F-theory compactifications},  J. Phys. {\bf  A35 } (2002) 
965, {\tt
  hep-th/0112005}.
\bibitem{37}
N. J. Hitchin, A. Karlhede, U. Lindstrom and M. Rocek, Commun. Math. Phys. 
{\bf 108 }(1987) 535.

 \bibitem{38} B. Acharya, {\em  On realizing $N=1$ super Yang-Mills in
  M-theory}, {\tt hep-th/0011089}\\
       B. Acharya, {\em  M theory, Joyce Orbifolds and Super Yang-Mills},
        Adv. Theor. Math. Phys. {\bf 3} (1999) 227, {\tt  hep-th/9812205}.
\bibitem{39} B. Acharya, {\em  M-theory, $G_2$-manifolds and four dimensional 
physics}, {\it  ICTP} Lecture  notes, Spring school on superstrings and 
related matters, March, 2002.
\bibitem{40}
J.A. Harvey, G. Moore, { \em Superpotentials and Membrane Instantons},
 {\tt  hep-th/9907026}.
\bibitem{41}
S. Kachru, J. McGreevy, {\em M-theory on Manifolds of  $G_2 $ Holonomy and
  Type IIA Orientifolds},   JHEP {\bf  0106} (2001) 027, {\tt  hep-th/0103223}.
   \bibitem{42} M.F. Atiyah and E. Witten, {\em  M-theory dynamics
  on a manifold of $G_2$ Holonomy},
   {\tt hep-th/0107177}.\\
       T. Friedmann, {\em  On the Quantum Moduli Space of M Theory 
Compactifications},    Nucl. Phys. {\bf B635} (2002) 384, {\tt  
hep-th/0203256}.
   \bibitem{43}
  M. Cvetic, G. Shiu, A.M. Uranga,
   {\em Chiral Type II Orientifold Constructions as M
  Theory on $G_2$ holonomy spaces}, {\tt hep-th/0111179}.

   \bibitem{44} E. Witten, {\em Anomaly Cancellation On Manifolds Of $G_2$
  Holonomy},
  { \tt hep-th/0108165}.

 \bibitem{45}  A. Belhaj, {\em  F-theory Duals of M-theory on $G_2$ Manifolds 
from Mirror Symmetry},  J. Phys. {\bf A36} (2003) 4191, {\tt  hep-th/0207208}.
\bibitem{46}
G. Curio, {\em  Superpotentials for M-theory on a $ G_2$  holonomy manifold
  and
  Triality symmetry},
 {\tt  hep-th/0212211}.
\bibitem{47}
 T. Friedmann, E. Witten, {\em Unification Scale, Proton Decay, And
  Manifolds
  Of $ G_2$  Holonomy}, {\tt
hep-th/0211269}.
 \end{thebibliography}
   \end{document}